# Galaxy Classification: A deep learning approach for classifying Sloan Digital Sky Survey images

Sarvesh Gharat,[1]⋆ Yogesh Dandawate[1]

[1]*Department of Electronics and Telecommunication, Vishwakarma Institute of Information Technology, Pune, India*

**ABSTRACT**

In recent decade, large scale sky surveys such as Sloan Digital Sky Survey (SDSS) has resulted in generation of tremendous amount of data. The classification of this enormous amount of data by astronomers is time consuming. To simplify this process, in 2007 a volunteer based citizen science project called "Galaxy Zoo" was introduced which has reduced the time for classification by a good extent. However, in this modern era of deep learning, automating this classification task is highly beneficial as it reduces the time for classification. Since last few years, many algorithms have been proposed which happen to do a phenomenal job in classifying galaxies into multiple classes. But all these algorithms tend to classify galaxies into less than 6 classes. However after considering the minute information which we know about galaxies, it is necessary to classify galaxies into more than 8 classes. In this study, a neural network model is proposed so as to classify SDSS data into 10 classes from an extended Hubble Tuning Fork. Great care is given to disk edge and disk face galaxies, distinguishing between a variety of substructures and minute features which are associated to each class. The proposed model consists of convolution layers to extract features making this method fully automatic. The achieved test accuracy is 84.73 % which happens to be promising after considering such minute details in classes. Along with convolution layers, the proposed model has 3 more layers responsible for classification which makes the algorithm consume less time.

## 1 INTRODUCTION

There are more than 2 trillion galaxies in our Universe (Conselice et al. 2016) (Smith 2016). Those observed so far exhibit a variety of morphologies through cosmic time, showing spiral, elliptical or irregular shapes and sub-structural features such as bars and clumps. Morphology of galaxy relates to its internal properties such as radio emission, star forming activity (Bell et al. 2003) (Kennicutt Jr 1998) and can reveal some insights about it's evolutionary history, including mergers (Mihos & Hernquist 1995) and interaction with environment (Sol Alonso et al. 2006).

The importance of morphological classification was first given by Hubble in 1926 (Hubble 1926) (Oswalt & Gilmore 2013) (Hernández-Toledo et al. 2008) as a descriptive system. In general, the classification was visually done by professional astronomers. However, this method of mannual classification has different pros and cons e.g. professional astronomers can classify images with high accuracy, but can only analyse limited amount of data.However, nowadays lots of algorithms have been proposed which extracts different features like color, shape, brightness profile, concentration index, etc. Compared to past methodologies, the usage of automatic feature-extraction algorithms can significantly reduce the time spent on analyzing galaxy images. These techniques include parametric and non parametric fitting. In parametric fitting 2D analytical functions are fitted on galaxy image. During the fitting routine the assumed mathematical model is convolved with the Point Spread Function (PSF) to account for atmospheric effects. Similarly in non parametric methods, analysis of light distribution in Petrosian radius is done (Tarsitano et al. 2018). Recently, a decade ago, a citizen science project was launched which has been great success in classifying large volumes of data (Simmons et al. 2016). However the speed of actual classification has been decreased by smaller extent. In today's era, there are large number of surveys which are expected to yield a large volumes of data. This large volume of data would need a large number of volunteers for doing classification. Therefore research nowadays is seeing the rise of automated algorithms for galaxy classification.

Recent advancement in machine learning and neural networks has given significant results. With that, the amount of knowledge we have about galaxies have also increased by a larger extent. Hence, it is must to classify galaxies in more classes than that of which traditional algorithms do. (Tarsitano et al. 2021) gave a detailed overview on different methods used in classification. The authors have also proposed a novel work in their paper. In this study, we propose a automated system to classify galaxies into multiple classes. The proposed algorithm uses convolution neural networks. Convolution layer is responsible for extracting features from images. Each convolution kernel extracts multiple features from the whole input plane (Liu 2018). The extracted features are further treated as an input to classifier which classifies galaxies into 10 classes. More information on why it is important to classify galaxies considering such deep features in given in Section 2

⋆ E-mail: sarveshgharat19@gmail.com

The processed data used in this study is collected from a python package named "AstroNN" (Leung & Bovy 2019) which has SDSS data in it. SDSS which stands for Sloan Sky Digital Survey is one of the most resolved survey which has been active since last 2 decades. After completing one decade of observation, SDSS had collected enough data which created a new 3D map of massive galaxies and distant black holes (Leung & Bovy sds).

## 2 RELATED WORK

Different classification techniques use different methodologies. Some techniques include manual feature extraction while some include automatic feature extraction from galaxy images or photometric data from galaxy images.

(Walmsley et al. 2020) proposed bayesian convolution neural network. The proposed technique is similar to VGG16 (Visual Geometry Group). VGG16 is a CNN based architecture which consists of 16 convolution layers (Simonyan & Zisserman 2014) developed by Visual Geometry Group Lab of Oxford University has 3 × 3 kernel of different sizes. The proposed technique by Walmsley et. al uses active learning which reduces the training data by 35% to 60%. The authors classify galaxies based on the bar present in it into barred and unbarred class. To do this, the authors make use of two variables namely $N_{bar}$ and $k_{featured}$. For galaxies having bar, they consider $N_{bar} \geq 10$. Similarly for unbarred, $N_{bar} < 10$ and $k_{featured} < 10$. The authors have almost used 3,04,122 images for this study.

(Mittal et al. 2020) have proposed a deep convolutional neural network to classify SDSS data in 3 classes i.e spiral, elliptical and irregular galaxies. The data used in this study is less i.e 4614 images, hence the authors have used data augmentation. The achieved accuracy which in 98% is the best accuracy attained by any model till now, however it is restricted only for 3 classes.

(Eassa et al. 2021) have focused on raw brightness level of galaxies. Based on brightness level and euclidean distance from centre, the proposed technique classifies the data into spherical, elliptical and irregular classes. Here, the authors have initially subtracted the background brightness due to stars and other objects so as to get raw brightness level from the galaxy. This is done by deciding a particular threshold and substracting all the values from that threshold. Then on using techniques like k means clustering, the galaxies are classified into 3 classes. The authors have used 1000 images to test and have achieved accuracy of 97.2%.

(Jiménez et al. 2020) have used auto-encoder for extracting features from images. The data used in this study is collected from GZ1(Galaxy Zoo 1) (Lintott et al. 2008) dataset and consists of more than 650k data images. Out of 650k images, almost 41k of images are classified by professional astronomers whereas remaining data is classified by amateur astronomers.

(Bom et al. 2021) have a proposed a RCNN model which allows pixel level segmentation. The study done in this work also focuses on detailed analysis of different learning techniques like transfer learning, differential learning and data augmentation. However, the number of classes in which the algorithm classifies the data are just 2 i.e. spiral and elliptical.

(Cavanagh et al. 2021) have proposed 4 different models for classifying data into 2 (elliptical and spiral), 3 (elliptical, spiral and lenticular)and 4 (elliptical, spiral, lenticular and others) classes respectively. Due to limitation of data, the authors have rotated and flipped the images to generate additional data images. The achieved accuracy after classifying Galaxy Zoo 2 (Willett et al. 2013) data is 92%, 82% and 77% for 2, 3 and 4 classes respectively.

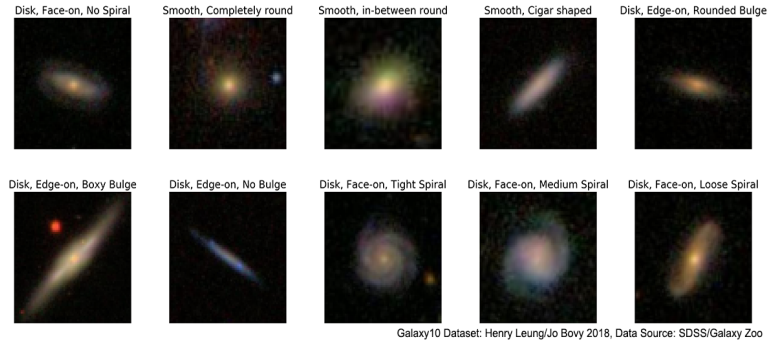

**Figure 1.** Images of each class in the dataset

All this studies have contributed to a good extent in classification of SDSS images. In recent days the amount of theoretical knowledge we have is immense, hence it is must to classify galaxies into more classes. However there's no algorithm to our knowledge which can classify this data into more than 8 classes with accuracy more than 75%. This study focuses on a deep learning approach which classifies the data in 10 classes which will solve this problem by a greater extent. The major reason behind classifying galaxies into 8 classes or more is the varying properties of galaxies in each and every class. Having classified after considering every minute feature, it is easy to study on particular type of galaxy in deep. eg: A radio loud galaxy is generally elliptical in shape. At the same time the elliptical galaxies can be divided based on radion jets into fr1 and fr2. Similarly, there are many such unique properties which are found in galaxies with particular morphology (Jiang et al. 2013) (Ferguson & Binggeli 1994) (De Vaucouleurs 1959) (Kormendy et al. 2009) (De Paz et al. 2003) (Laurikainen et al. 2005) (Graham 2019).

## 3 METHODOLOGY

This section focuses on the methodology proposed in the study.

### 3.1 Data Collection and Preprocessing

The SDSS catalogue considered in this study contains 21785 visually classified samples from Galaxy Zoo data. All the images in this study are three band (*r*, *g* and *i* band) colored images which are classified into 10 classes. As Galaxy Zoo depends on volunteers for classification of data, there are chances that volunteers may make mistakes during classification, hence we only take the images for which more than 55% of people have selected a particular class. The original image in this dataset were of 424 x 424 pixels, which are further cropped to 207 x 207 pixels and then downscaled using bilinear interpolation (Gribbon & Bailey 2004) (Rukundo & Maharaj 2014) to 69 x 69 pixels.

We have used a python package named astroNN (Leung & Bovy 2019) to import the dataset for our study.

### 3.2 Data Distribution

In this study, data has been divide into 10 classes .

Figure 1 shows us examples of data belonging to each class and Table 1 gives us the number of images present in each class.

**Table 1.** Distribution of data

| Class | Name | Number of Images |
|---|---|---|
| Class 0 | Disk, Face-on, No Spiral | 3461 |
| Class 1 | Smooth, Completely round | 6997 |
| Class 2 | Smooth, in-between round | 6992 |
| Class 3 | Smooth, Cigar shaped | 394 |
| Class 4 | Disk, Edge-on, Rounded Bulge | 1534 |
| Class 5 | Disk, Edge-on, Boxy Bulge | 17 |
| Class 6 | Disk, Edge-on, No Bulge | 589 |
| Class 7 | Disk, Face-on, Tight Spiral | 1121 |
| Class 8 | Disk, Face-on, Medium Spiral | 906 |
| Class 9 | Disk, Face-on, Loose Spiral | 519 |
| | Total | 21785 |

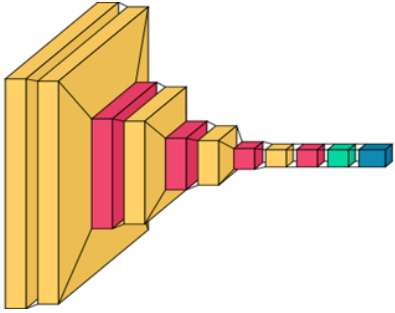

**Figure 2.** Feature Extracting Model

## 3.3 Model Architecture

The proposed Neural Network model is divided into 2 parts.

### 3.3.1 Feature Extraction Model

In this model, we have used convolution layers to extract features. Convolution Layers are meant to extract high level features from the image provided in input plane (Gu et al. 2018) (Alzubaidi, Zhang, Humaidi et al. Alzubaidi et al.) (Albawi et al. 2017). Convolution Layers apply the filter to input image to create a feature map. Each 2D convolution layer takes $n$ x $n$ array and applies $k$ filters with help of $m$ x $m$ kernel so as to extract $k$ features. Here, $n$ x $n$ are dimensions of input image (individually $r$, $g$, $b$ bands), $m$ x $m$ are the dimensions of convolution kernel and k is the number of filters (Gu et al. 2018) (LeCun et al. 1989) (LeCun et al. 1998). With convolution layers, we use pooling layers to downsample the feature map. In this work, we have used MaxPooling which helps to preserve maximal elements and reduces the noise (Christlein et al. 2019) (Gholamalinezhad & Khosravi 2020). Each convolution layer used in this model has "Relu" as an activation function. Relu is an activation function which happens to be 0 for all negative inputs and same as input for all the positive values. This is done to increase the non linearity of the last component in each convolution layer. After extracting features with help of convolution and pooling layer, all the feature maps are flattened into a 1D array.

In Figure 2, convolution layer is represented by yellow color, pooling layer by red color, dropout by green color and flattening layer by blue color respectively.

As it is evident from Figure 2, the feature extracting model has multiple convolution blocks starting with convolution layer, followed by a block of alternating convolution layer and MaxPooling layer.

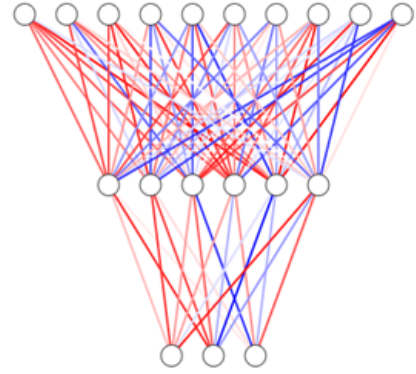

**Figure 3.** Classifier Model

### 3.3.2 Classifier Model

The output of feature extracting model acts as an input to the classifier.

The classifier has 2 dense layers before the output layer as seen in Figure 3. The first dense layer in the classifier has 64 neurons, whereas the second layer has 32 neurons. As the number of output classes are 10, the number of neurons in the output layer is 10. The activation function used in first 2 dense layers is "Relu", in the output layer we use "Softmax" as a activation function. This is done so that the output of the model will be the probability of image belonging to particular class (Agarap 2018) (Feng & Lu 2019) (Nwankpa et al. 2018). To avoid overfitting, a regularisation technique named dropout is used in both densed layers. As dropout of 0.2 is used, 20% of neurons act as dead neurons (Srivastava et al. 2014) (Cai et al. 2019) (Srinivas & Babu 2016).

## 3.4 Training and Validation

The proposed model is constructed using keras (Chollet et al. 2015). The training of model is conducted on Google Collaboratory, making use of NVIDIA K80 GPU (Carneiro et al. 2018). The dataset in this study is divided in train, test and validation dataset in the ratio of 70:15:15 i.e. 70% of data is used for training (15249 images), 15% of data is used for testing (3268 images) and remaining 15% of data is used in validation dataset (3268 images). Validation dataset which is independent of testing dataset is used for hyperparameter tuning so as to avoid any biasing in choice of hyperparameters. Thus, when the network is completely trained, evaluation is done on completely unseen test dataset.

Training was conducted over maximum of 50 epochs. For efficient training, we use early stopping so that the training ends once there is no appreciable decrease in validation loss. Early stopping helps to avoid unnecessary computation once validation accuracy has reached it's peak and also avoids overfitting (Montavon et al. 2012) (Ying 2019) (Song et al. 2019) (Caruana et al. 2001).

In Figure 4, blue colour represents training loss and accuracy whereas red colour represents validation loss and accuracy.

As seen in Figure 4, during initial epochs the gradient of loss is high but at later stage it decreases. Here's when early stopping helps. If there was no early stopping the loss would have further increase, making the accuracy to decrease. The validation loss and accuracy as seen in Figure 4 represents a metric corresponding to repeated evaluation of networks.

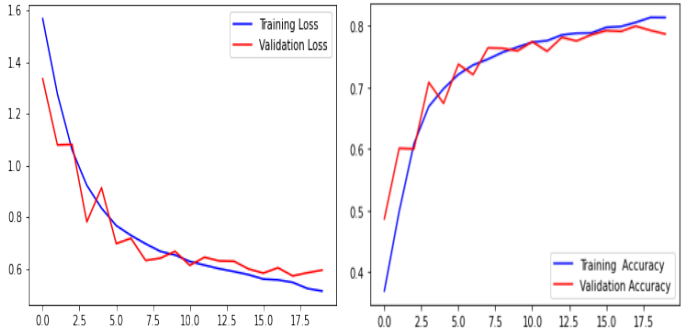

**Figure 4.** Graph of Accuracy and Loss against number of epochs

**Table 2.** Confusion Matrix for testing dataset

|   | 0 | 1 | 2 | 3 | 4 | 5 | 6 | 7 | 8 | 9 |
|---|---|---|---|---|---|---|---|---|---|---|
| 0 | 342 | 52 | 60 | 3 | 6 | 0 | 1 | 29 | 11 | 10 |
| 1 | 15 | 987 | 11 | 0 | 0 | 0 | 0 | 4 | 0 | 0 |
| 2 | 30 | 60 | 850 | 2 | 1 | 0 | 0 | 0 | 0 | 0 |
| 3 | 2 | 0 | 4 | 53 | 4 | 0 | 7 | 0 | 0 | 2 |
| 4 | 16 | 0 | 7 | 3 | 206 | 1 | 15 | 0 | 0 | 1 |
| 5 | 0 | 0 | 0 | 0 | 1 | 3 | 0 | 0 | 0 | 0 |
| 6 | 0 | 0 | 0 | 3 | 20 | 0 | 67 | 0 | 0 | 1 |
| 7 | 22 | 5 | 2 | 0 | 2 | 0 | 0 | 101 | 10 | 0 |
| 8 | 20 | 1 | 2 | 0 | 0 | 0 | 0 | 11 | 99 | 9 |
| 9 | 10 | 0 | 3 | 1 | 4 | 0 | 2 | 3 | 16 | 61 |

## 4 RESULTS AND DISCUSSION

Almost all the previous studies done in classification of galaxies are done to classify galaxies into at most 5 classes (Walmsley et al. 2020) (Mittal et al. 2020) (Jiménez et al. 2020) (Eassa et al. 2021) (Bom et al. 2021) (Cavanagh et al. 2021). But considering the development in theoretical aspects, we found that there was need of classifying it into more classes. Reason for it is given in section 2.

In this study we define a convolutional neural network as mentioned in section 3 which has been trained on SDSS dataset to classify galaxies in 10 classes.

In Table 2, verticaly we have the number of images predicted by algorithm belonging to particular class. On the other hand, on horizotal axes we have number of images that actually belong to particular class. We know that the overall accuracy of the model is given by

$$\text{Accuracy} = \frac{\sum_{i=0}^{9} a_{ii}}{\sum_{i=0}^{9} \sum_{j=0}^{9} a_{ij}} \times 100$$

In our case it will be

$$\text{Accuracy} = \frac{342 + 987 + \cdots + 61}{342 + 52 + 60 + \cdots + 4 + 16 + 61} \times 100 = 84.73\%$$

From confusion matrix we can see that 172 images (33.4 %) are misclassified as class 0, this is because of similarity which class 0 i.e Disk Face on and No spiral has with other classes. Also class 7, 8, 9 just vary in the structural wound they have. This is commented after visually looking images from all the classes.

**Table 3.** Precision, Recall and F1 score

| Class | Precision | Recall | F1 score |
|---|---|---|---|
| Class 0 | 0.748 | 0.665 | 0.704 |
| Class 1 | 0.893 | 0.971 | 0.930 |
| Class 2 | 0.905 | 0.901 | 0.902 |
| Class 3 | 0.815 | 0.736 | 0.773 |
| Class 4 | 0.844 | 0.827 | 0.835 |
| Class 5 | 0.750 | 0.750 | 0.750 |
| Class 6 | 0.728 | 0.736 | 0.731 |
| Class 7 | 0.682 | 0.711 | 0.696 |
| Class 8 | 0.728 | 0.697 | 0.712 |
| Class 9 | 0.726 | 0.610 | 0.662 |
| Average | 0.782 | 0.760 | 0.769 |

Another mean of testing the model is to calculate overall F1 score.

$$F1 = \frac{2PR}{P + R}$$

With reference to Table 3, Precision tells about the false positive rate i.e precision of class 0 will tell about the images misclassified as class 0. Similarly, Recall tells about the false negative rate. The value of Precision and Recall ranges from 0 to 1. The F1 score for any model is defined as the harmonic mean of Precision and Recall. It is calculated so as to balance these two parameters.

In each class reported in the table, the higher are the values of the three parameters, the better is the model. The last three classes have very similar morphology and they show similar accuracies, hence the probability of misclassification is higher. Also, as discussed earlier, lots of images (172 images) are misclassified as class 0, so the lower F1 score of classes 0, 7, 8 and 9 contribute to decrease the average F1 score of the model.

### 4.1 Conclusion

In this study, we propose a convolutional neural network to classify galaxies in 10 classes. This is one of the initial work wherein the galaxies are classified in 10 classes by considering such minute details. This detailed classification happens to be of need after considering the theoretical knowledge we have. This was initially done by professional astronomers but due to large amount of data we have, it was impossible for them to continue. Further, different citizen scientists were trained to do this task, but in modern era the data we have is huge in number compared to that of available volunteers. Hence, this algorithm happens to solve this problem. Also, the time taken by algorithm to classify large volume of dataset is less than 10 minutes which is another advantage of using the automated algorithms over manual classification. The proposed algorithm gives accuracy of 84.5 % which is good after considering such minute details in classification.

## CONFLICT OF INTEREST

The authors declare that they have no conflict of interest.

## DATA AVAILABILITY

All the data used in this study is publicly available on astronn's official website. The code used for the proposed model will be made available after 2 years of publishing on request.

This paper has been typeset from a T<sub>E</sub>X/LAT<sub>E</sub>X file prepared by the author.